\documentclass[preprint,preprintnumbers,amsmath,amssymb]{revtex4}

\usepackage{graphicx}
\usepackage{dcolumn}
\usepackage{bm}
\begin{document}

\title{ Persistent Current  for  a genus g=2 structure}

\author{D. Schmeltzer }

\affiliation{Physics Department, City College of the City University of New York \\
New York, New York 10031}

\date{\today}
\begin{abstract}
We report the persistent current in two coupled rings which  form a character ``8'' structure. We showed that the magnetic fluxes modify the global phase of the electronic wave function for multiple connected geometry formed by the coupled rings. We obtained an exact solution for the persistent current and investigated the exact solution numerically. For two large coupled rings with equal fluxes, we found that the persistent current in the two coupled rings is in fact equal to that in a single ring.  For opposite fluxes the  energy has a chaotic structure. For both cases the periodicity is  $h/e$. This results are obtained within an  extension of  Dirac's  second class constraints. This theory can be tested in the ballistic regime.
\end{abstract}

\pacs{73.23.Ra }
\keywords{Dirac's Constraints, Persistent Current, High Genus Material}
\maketitle
\textbf{I Introduction}

\vspace{0.2 in}

In Quantum Mechanics, the wave function is path dependent and is sensitive to the presence of a vector potential caused by an external magnetic flux. In a closed geometrical structure, such as a ring the wave function is changed by a measurable phase [1], causing all the physical properties to become periodic [2]. When a mesoscopic ring of normal metal is pierced by a magnetic flux $\Phi$ [2], the boundary conditions are modified, leading to a famous theorem of periodic properties with the flux period $\Phi_0 = h / e$ and to a remarkable phenomena [3] of a non-dissipative \textit{persistent current} [3-7].

One way to classify the closed geometrical structure is by using the number of $holes$ formed on the closed geometrical structure. For a $closed$ $surface$, the number of holes formed thereon is often referred to as a genus number $g$ [9, 11]. For example, a genus number $g=1$ describes an Aharonov-Bohm ring geometry, while a genus number $g = 2$ describes two rings perfectly glued at one point to form a character ``8'' structure.

In this paper, we report the $first$ exact solution  for multiple connected geometries, such as a geometry with two  holes  \textbf{two rings perfectly glued at one point to form a character ``8'' structure.}
 or connected rings). The geometry modifies the $global$ properties of the wave function, and the presence of magnetic fluxes generates persistent currents with complicated periods. We present an exact analytical solution for the eigenvalues and compute the persistent current for two coupled rings with a character ``8'' structure for two different fluxes. We solve the problem by modeling the $gluing$ of the two rings using $Fermionic$ constraints  with $anti$-$commuting$ $Lagrange$ multipliers, which can be viewed as a resonant impurity strongly coupled to the two rings .

The analytical results are investigated numerically. When the two fluxes on both rings are the same, we find a simple relation between the single ring ($g=1$) current, $I^{(g=1)}(flux;N)$, and the double ring ($g=2$) current, $I^{(g=2)}(flux;N)$. At $T=0$, we define $ I^{(g=2)}(flux;N) = r(N) I^{(g=1)}(flux;N)$, where $r(N)$ is a ratio between the two currents. The ratio $r(N)$ is a function of the number of sites $N$ and obeys $r(N) \rightarrow 1$ for $N \rightarrow \infty$.

The plan of this paper is as followings  in chapter $II$ we present the exact analytical results for the two rings perfectly glued at one point to form a character ``8'' structure.
In chapter $III$ we present the numerical results for the two coupled rings for equal and opposite flues.
Chapter $IV$ is devoted to the discussions:
  We show that the method presented can be extended to two rings with a finite with.
  
   Using  the one dimensional Persistent current  (single channel) given in ref.9  in the presence of  a $2K_{F}$ impurity scattering  given in ref.9  we compute the current to  the  multichannel case. We show that the  theory developed for the  $Ballistic$ regime might be able  to  explain the  experimental results for the $GaAs/GaAlAs$ coupled rings [13].

We also present an generalization  of the  Dirac's constraint  method  for  a genus $g>2$ multi coupled rings.

\textbf{II Exact Solution for two rings perfectly glued at one point to form a character ``8'' structure}

Dirac has shown  [14 ] that for the  second class constraints the $Poisson$ $brackets$  are replaced by the $Dirac$ [14] $brackets$.
 For an $even$ number of  $constraints$  $ q_{r}$ with $non$ zero $Poisson$
$brackets$, the equations of motions are governed by  the Dirac [14] brackets which   replace 
the  Poisson bracket $\{A,B\}$ by ,$\{A,B\}_{D}=\{A,B\}-\sum_{r,r^{\prime }}\{
A,q_{r}\} c_{r r^{\prime }}\{q_{r^{\prime }},B\}$ .
The  matrix $c_{r r^{\prime }}$  is given in terms of the constraints  $q_{r}$ , $\sum\limits_{r^{\prime
}}c_{r r^{\prime }}\{q_{r^{\prime },}q_{s}\}=\delta _{r,s}$.
To obtain a Bosonic theory  one replaces the Poisson bracket by the commutators, $\{,\}=i\hbar [,]$.

\textit{We propose  that for  the second class Fermionic constraints the following modification.} 
 Given two Ferminonic constraints $Q$, $Q^{\dagger }$ which obey
$non$-$zero$ $anticommutation$  relations, 
$\left[ Q,Q^{\dagger }\right] _{+}\equiv
QQ^{\dagger }+Q^{\dagger }Q\neq 0$. 
\textit{We find    that the Dirac bracket is replaced  for a Fermionic operator  $\widehat{O}_{F}$ and the  hamiltonian $H_{0}$   by:}

$\left[ \widehat{O},H_{0}\right] _{D} \equiv \left[ \widehat{O},H_{0}\right] -%
\left[ \widehat{O},Q^{\dagger }\right] _{+}\left( \left[ Q^{\dagger },Q%
\right] \right) ^{-1}  
\left[ Q,H_{0}\right] -\left[ \widehat{O},Q\right] _{+}\left( \left[
Q^{\dagger },Q\right] _{+}\right) ^{-1}\left[ Q,H_{0}\right] $

 Therefore   \textit{ the new  Heisenberg equation
 for any  fermionic operator $\widehat{O}_{F}$ will be given by:} 
$i\hbar \frac{d\widehat{O}_{F}}{dt}=\left[ \widehat{O}_{F},H_{0}\right] _{D}$.
For the remaining part we present the derivation and applications of this new result.

We consider the Hamiltonian $H_{0}$ for two spinless Fermionic rings in the $absence$ of a magnetic flux. The rings obey periodic boundary conditions. For each ring, the point $x$ is identified with the point $x+L$. The two coupled rings with the character ``8'' structure (i.e. $g=2$) are obtained by identifying the middle point $x = L/2$ of the first ring with point $x=0$ of the second ring, i.e. $C_1(L/2) = C_2(0)$ and $C^+_1(L/2) = C^+_2(0)$. This identification is equivalent to two $Fermionic$ $constraints$, $Q \equiv C_1(L/2) - C_2(0)$ and $Q^+ \equiv C^+_1(L/2) - C^+_2(0)$. Since the constraints are Fermionic, they can be enforced by using $anti$-$commuting$ Lagrange multipliers, $\mu^+$ and $\mu$. Following ref. [14], we introduce the Hamiltonian with the constraints,$H_{T} = H + \mu^+ Q + Q^+ \mu$.
The unusual physical meaning of the anti-commuting Lagrange multipliers can be viewed as a Fermionic impurity [14], which mediates the hopping of the electrons between the two rings. This method is simpler in comparison with the  method based on matching boundary conditions for the wave function explained in the discussions paragraph.
The two rings of length $L$ are threaded by a magnetic flux $\Phi_\alpha$, where $\alpha = 1, 2$ (for each ring). In order to observe the changes of the constraints in the presence of the external flux, we perform the following steps. In the $absence$ of the $external$ $flux$ $\Phi_\alpha$, the annihilation and creation fermion operators obey periodic boundary conditions $C_\alpha(x) = C_\alpha (x + L)$ and $C^+_\alpha(x) = C^+_\alpha (x + L)$, where $\alpha = 1, 2$. The genus $g=2$ is implemented by the Fermionic constraints $Q = C_1(L/2) - C_2(0)$ and $Q^+ = C^+_1(L/2) - C^+_2(0)$, and Hamiltonian $H_0 = - t \sum_{\alpha=1}^2 \sum_{x = 0}^{(N_{s} - 1) a} [C_\alpha^+(x) C_\alpha(x + a) + h.c.]$. The length of each ring is $L = N_{s} a$, where $N_{s}$ is the number of sites and $a$ is the lattice spacing. When the $external$ $magnetic$ $flux$ $\Phi_\alpha$ is appled the Hamiltonian $H_0$ is replaced by $H$. The Hamiltonian $H$ is obtained
by the transformation $C_\alpha(x) \rightarrow \exp[i \frac{e}{\hbar c} \int_0^x A (x';\alpha) d x' ] C_\alpha(x) \equiv \psi_\alpha(x)$ and $C^+_\alpha(x) \rightarrow C^+_\alpha(x) \exp[ - i \frac{e}{\hbar c} \int_0^x A (x';\alpha) d x' ] \equiv \psi^+_\alpha(x)$. Here $A(x;\alpha)$ is the $tangential$ component of vector potential on each ring. The relation between the flux and the vector potential on each ring is $\frac{e}{\hbar c} \int_0^L A(x;\alpha) dx  =\varphi_\alpha $.

The flux $\Phi_\alpha$ on each ring $\alpha=1, 2$ gives rise to a change in the boundary conditions, $\psi_\alpha(x + N_{s} a)= \psi_\alpha(x) e^{i \varphi_\alpha}$ and $\psi^+_\alpha(x + N_{s} a) = \psi^+_\alpha(x) e^{- i \varphi_\alpha}$, where $\varphi_\alpha = 2 \pi (\frac{e \Phi_\alpha}{h c} ) = 2 \pi \frac{\Phi_\alpha}{\Phi_0} \equiv 2 \pi \hat{\varphi}_\alpha$. This boundary condition gives rise to a normal mode expression for each ring, $\psi_\alpha(x) = \frac{1}{\sqrt{N}} \sum_{n=0}^{N_{s}-1} e^{i K(n, \varphi_\alpha) \cdot x} \psi_\alpha(n)$ and a similar expression for $\psi^+_\alpha(x)$. The ``momentum'' $K(n, \varphi_\alpha)$ is given by,
$K(n, \varphi_\alpha ) = \frac{2 \pi}{N_{s} a} (n + \hat{\varphi}_\alpha)$
where $n = 0, 1, \ldots, N-1$ are integers with $N = N_{s}$, and $\varphi_\alpha = 2 \pi \hat{\varphi}_\alpha$. In the momentum space, the Fermionic operators $\psi_\alpha(n)$ and $\psi^+_\beta(m)$ obey anti-commutation relations, $[ \psi_\alpha(n), \psi^+_\beta (m) ]_+ = \delta_{\alpha, \beta} \delta_{n,m}$. The Hamiltonian for the two rings in the presence the external flux takes the form,
\begin{equation} \label{eq:hamiltonian}
H = - t \sum_{\alpha = 1, 2} \sum_{x = 0}^{(N_{s}-1) a} \psi^+_\alpha (x) \psi_\alpha(x + a) + h.c. = \sum_{\alpha = 1, 2} \sum_{n = 0}^{N_{s}-1} \epsilon(n, \hat{\varphi}_\alpha) \psi^+_\alpha(n) \psi_\alpha (n)
\end{equation}
where $\epsilon(n, \varphi_\alpha) = - 2 t \cos [ \frac{2 \pi}{N} (n + \hat{\varphi}_\alpha)]$ are the eigenvalues for each ring. The Hamiltonian in eq. \ref{eq:hamiltonian} has to be solved $together$ with the $transformed$ $constraints$ ,$Q = \psi_1(\frac{L}{2}) e^{-i\frac{e}{\hbar c} \int_0^{\frac{L}{2}} A(x;\alpha=1) dx}  - \psi_2(0)$ and $Q^+ = \psi^+_1(\frac{L}{2}) e^{ i\frac{e}{\hbar c} \int_0^{\frac{L}{2}} A(x;\alpha=1) dx } - \psi^+_2(0)$.

The wave function for the genus $g=2$ problem is given by the eigenstate $ | \chi \rangle $ of the Hamiltonian in eq. \ref{eq:hamiltonian}, which in addition satisfies the equations $Q | \chi \rangle = 0$ and $Q^+ | \chi \rangle = 0$. The constraint conditions are implemented with the help of the $anti-commuting$ $Lagrange$ $multipliers$ $\mu$ and $\mu^+$. The Hamiltonian $H_{T}$ with the constraints takes the form,

$H_{T} = H + \mu^+ Q + Q^+ \mu$

The Lagrange multiplier are determined by the condition that the constraints are satisfied at any time. Therefore, the time derivative satisfies the equation, $\dot{Q} | \chi \rangle = \dot{Q}^+ | \chi \rangle = 0$ at any time. We will use the notations, $[A, B]_+ \equiv A B + B A$ and $ [A, B] = A B - B A$.
The Heisenberg equation of motion for the constraint $Q$ is,
\begin{eqnarray} \nonumber
i \hbar \dot{Q}&=& [Q, H_{T}] = [Q, H] + [Q, \mu^+ Q + Q^+ \mu ] \\ \nonumber
               &=& [Q, H] + [Q, \mu^+]_+ Q - \mu^+ [ Q, Q]_+ + [Q, Q^+]_+ \mu - Q^+ [ Q, \mu]_+ \\ \label{eq:heom}
               &=& [Q, H] + [Q, Q^+]_+ \mu
\end{eqnarray}
The rest of the anti-commutators in eq. \ref{eq:heom} vanishes. The anti-commuting Lagrange multipliers satisfy, $[Q, \mu^+]_+ = [Q, \mu]_+ = [Q^+, \mu^+]_+ = [Q^+, \mu]_+ = 0$. Since the constraints are fermionic, we obtain that they obey $[Q, Q^+]_+ = [Q^+, Q]_+ = 2$. Therefore, the constraints are second class constraints [14]. From the condition $\dot{Q} | \chi \rangle = 0$ and eq. \ref{eq:heom}, we determine the Lagrange multiplier field $\mu$,

$\mu = - [Q^+, Q]_+^{-1} [Q, H] = - \frac{1}{2} [Q, H]$. 

The field 
$\mu^+$ is obtained from the equation $\dot{Q}^+| \chi \rangle = 0$,

$\mu^+ = [ Q, Q^+]_+^{-1} [Q^+, H] = \frac{1}{2} [Q^+, H]$.

The Hamiltonian $H_{T}$ with the constraints and the $Lagrange$ multipliers are used to compute the $Heisenberg$ $equation$ of $motion$ for any $Fermionic$ $operator$, $\hat{O}$. (The Lagrange multipliers anti-commute with any Fermionic operator, i.e. $[\hat{O}, \mu]_+ = [\hat{O}, \mu^+]_+ = 0$.)
\begin{eqnarray} \nonumber
i \hbar \frac{d \hat{O}}{d t} = [ \hat{O}, H_{T}] &=& [\hat{O}, H] + [\hat{O}, \mu^+ Q] + [ \hat{O}, Q^+] \mu \\ \nonumber
&=& [\hat{O}, H] + [\hat{O}, \mu^+ ]_+ Q - \mu^+ [ \hat{O}, Q]_+ + [\hat{O}, Q^+]_+ \mu - Q^+ [ \hat{O}, \mu]_+ \\ \label{eq:oeom}
&=& [\hat{O}, H] - [\hat{O}, Q]_+ \mu^+ - [\hat{O}, Q^+] \mu
\end{eqnarray}
We substitute in eq. $3$ the solutions for the Lagrange multiplier fields and obtain a $new$ $equation$ of $motion$ with a $new$ $commutator$, which resemble the classical Dirac brackets [14].
\begin{eqnarray} \nonumber
i \hbar \frac{d \hat{O}}{d t} = [ \hat{O}, H_{T}] &=& [\hat{O}, H] - [\hat{O}, Q^+]_+ ([Q^+, Q]_+)^{-1} [Q, H] - [\hat{O}, Q]_+ ([Q, Q^+]_+)^{-1} [Q^+, H] \\ \label{eq:dirac}
&\equiv& [\hat{O}, H ]_D
\end{eqnarray}
Eq. \ref{eq:dirac} shows that the Heisenberg equation of motion is governed by a $new$ $commutator$, $[\hat{O}, H ]_D$. The equations $Q | \chi \rangle = 0$ and $Q^+ | \chi \rangle = 0$ are  $inconsistent$ with  $[Q,Q^+]| \chi \rangle \neq 0$. The new $commutator$ $resolves$ the $inconsistency$ problem,$[Q,Q^+]_{D}| \chi \rangle =0$! We will use this new commutator to compute the Heisenberg equations of motion for the creation and annihilation Fermionic operators $\psi_\alpha(x,t)$ and $\psi_\alpha^+(x,t)$, where $\alpha = 1, 2$.
\begin{eqnarray} \nonumber
i \hbar \dot{\psi}_\alpha(x) &=& [ \psi_\alpha(x), H]_D = [ \psi_\alpha(x), H] - \frac{1}{2}[ \psi_\alpha(x), Q^+]_+ [Q,H] \\ \nonumber
&=& - t [ \psi_\alpha(x + a)+ \psi_\alpha(x-a)] - \frac{1}{2}[ \delta_{\alpha, 1} \delta_{x, L/2} e^{i \varphi_1} - \delta_{\alpha, 2} \delta_{x, 0}] \\ \label{eq:4}
& & \cdot (-t) \{ e^{-i\varphi_1} [ \psi_1(\frac{L}{2} + a) + \psi_1(\frac{L}{2}-a) ] +
e^{-i\varphi_2} [ \psi_2(\frac{L}{2} + a) + \psi_2(\frac{L}{2}-a) ] \}
\end{eqnarray}

The ground state wave function is obtained from the one electron state, $|\chi> = \sum_{\alpha = 1, 2} \sum_{x = 0}^{(N_{s}-1) a} Z_\alpha(x) \psi_{\alpha}^+ (x) |0 >$, given in terms of the site amplitudes $Z_\alpha(x)$. Using a normal mode momentum expansion, $f_\alpha(n)$, i.e. $Z_\alpha(x) = \frac{1}{\sqrt{N}}\sum_{n=0}^{N-1} e^{i K(n, \hat{\varphi}_\alpha) x} f_{\alpha}(n)$, we find the following equations for the eigenvalues $\lambda$ and the amplitudes in the momentum space $f_{\alpha}(n)$,
\begin{equation} \label{eq:normal1}
(\lambda - \epsilon(\ell + \hat{\varphi}_1)) f_1(\ell) = - \frac{e^{i \pi \ell}}{2N} \sum_{n=0}^{N-1} \epsilon(n + \hat{\varphi}_1) e^{i \pi n} f_1(n) - \frac{1}{2N} \sum_{n=0}^{N-1} \epsilon(n + \hat{\varphi}_2) f_2(n)
\end{equation}
and
\begin{equation} \label{eq:normal2}
(\lambda - \epsilon(\ell + \hat{\varphi}_2)) f_2(\ell) = \frac{1}{2N} \sum_{n=0}^{N-1} \epsilon(n + \hat{\varphi}_2) f_2(n) + \frac{e^{i \pi \ell}}{2N} \sum_{n=0}^{N-1} \epsilon(n + \hat{\varphi}_1)e^{i \pi n} f_1(n)
\end{equation}

We diagonalize eqs. \ref{eq:normal1} and \ref{eq:normal2} by linear transformations,
$S_1 (\hat{\varphi}_1, \lambda) = -\sum_{\ell = 0}^{N-1} \epsilon(\ell+\hat{\varphi}_1) e^{i\pi\ell} f_1(\ell)$ and $S_2 (\hat{\varphi}_2, \lambda) = -\sum_{\ell = 0}^{N-1} \epsilon(\ell+\hat{\varphi}_2) f_2(\ell)$.
As a result, we obtain the equation,
$\textbf{M} \left(%
\begin{array}{c}
  S_1 \\
  S_2 \\
\end{array}
\right) = 0$,
where the matrix $\textbf{M}$ is given by $\textbf{M} = \left(%
\begin{array}{cc}
  -(1+\Delta^{(+)}_1) & \Delta^{(-)}_1 \\
       \Delta^{(-)}_2 & 1 - \Delta^{(+)}_2 \\
\end{array}%
\right)$.
Here, we define $\Delta^{(+)}_{\alpha}(\hat{\varphi}_\alpha, \lambda )\equiv\Delta_{\alpha}^{(even)} (\hat{\varphi}_\alpha, \lambda) + \Delta_{\alpha}^{(odd)} (\hat{\varphi}_\alpha, \lambda)$ and $\Delta^{(-)}_{\alpha}(\hat{\varphi}_\alpha, \lambda )\equiv\Delta_{\alpha}^{(even)} (\hat{\varphi}_\alpha, \lambda) - \Delta_{\alpha}^{(odd)} (\hat{\varphi}_\alpha, \lambda)$, with the $even$ and $odd$ representations given by, $\Delta_{\alpha}^{(even)} (\hat{\varphi}_\alpha, \lambda)=\frac{1}{2N} \sum_{m = 0}^{(N-2)/2} \frac{\epsilon( 2m + \hat{\varphi}_\alpha ) }{ \lambda - \epsilon(2 m + \hat{\varphi}_\alpha)}$ and $\Delta_{\alpha}^{(odd)} (\hat{\varphi}_\alpha, \lambda)=\frac{1}{2N} \sum_{m = 0}^{(N-2)/2} \frac{\epsilon( 2m + 1 + \hat{\varphi}_\alpha ) }{ \lambda - \epsilon(2 m + 1 + \hat{\varphi}_\alpha)}$. We compute $\det \textbf{M} = 0$ and obtain the $characteristic$ $polynomial$ which is used to determine the $eigenvalues$ $\lambda$.
\begin{equation}
2 [ \Delta_1^{(even)} (\hat{\varphi}_1, \lambda) \Delta_2^{(odd)} (\hat{\varphi}_2, \lambda) + \Delta_1^{(odd)} (\hat{\varphi}_1, \lambda) \Delta_2^{(even)} (\hat{\varphi}_2, \lambda)] + [\Delta^{(+)}_1 (\hat{\varphi}_1, \lambda) - \Delta^{(+)}_2 (\hat{\varphi}_2, \lambda)] = 1
\label{eq:secular}
\end{equation}
Eq. \ref{eq:secular} is our main result for the genus $g=2$ case. We observe that the matrix $M$ is $symmetric$ and the eigenvalues are real when the fluxes are equal, i.e. $\hat{\varphi}_1=\hat{\varphi}_2$, or opposite, i.e. $\hat{\varphi}_1=-\hat{\varphi}_2$. For other cases, the eigenvalues can have imaginary parts, thereby giving rise to non-conducting states.

\vspace{0.1 in}

\textbf{III Numerical Solution }

\vspace{0.1 in}

We have numerically solved the secular equation \ref{eq:secular}. To compute the current, we sum over the current carried by each eigenvalue $\lambda(\hat{\varphi}_1,\hat{\varphi}_2)$ using the grand-canonical ensemble. The current in each ring $\alpha=1, 2$ is given by,
$I^{(g=2)}_{\alpha}(\hat{\varphi}_1,\hat{\varphi}_2) = -\sum_{\lambda(\hat{\varphi}_1,\hat{\varphi}_2)}{\frac{d}{d\hat{\varphi}_\alpha}[ \lambda(\hat{\varphi}_1,\hat{\varphi}_2)}]F(\frac{(\lambda(\hat{\varphi}_1,\hat{\varphi}_2)-E_{fermi})}{K_{Boltzman}T})$
where $F(\frac{(\lambda(\hat{\varphi}_1,\hat{\varphi}_2)-E_{fermi})}{K_{Boltzman}T})$ is the Fermi Dirac function with the chemical potential $E_{fermi}$ and temperature $T$. The current is sensitive to the number of electrons being either even or odd. We use the grand-canonical ensemble and limit ourselves to a situation with even numbers of sites and a zero chemical potential, i.e. $E_{fermi}=0$ (which corresponds to the half-filled case). In order to have a perfect particle-hole symmetry, we will $restrict$ the analysis to the $special$ series for the $number$ of $sites$ being $N_{s}= 2, 6, 10, 14, 18, \ldots, 2m+2$, where $m = 0, 1, 2, 3 \ldots$. For this case, we find that, when the fluxes are the same in both rings, the current for $g=2$ has the same periodicity as that of a single ring, i.e. $I^{(g=2)}(\Phi + \Phi_0) = I^{(g=2)}(\Phi)$. At temperatures $T \leq 0.02$ Kelvin, the line shape of the current as a function of the flux is of a $sawtooth$ form (see figure $1b$). For other series $N_{s}\neq 2m+2$, the periodicity of the current is complicated. Using the experimental values given in the experiment [11], we estimate that the number of sites in our model should be in the range of $N_{s}= 50 \sim 150$, the $hopping$ constant should be $t= \frac{\hbar v_{fermi}}{2 a sin(K_{fermi}a)} \approx 0.01 $ eV  and the temperature in the experiment should be $T=0.02$ Kelvin. Using these units, we obtain that the persistent current is given in terms of a $dimensionless$ $current$, $I$ (see figure $1b$ and figure $1c$) with the actual current value, $I^{(g=2)}= I\times 0.92\times 10^{-4}$ Ampere.

\textit{a) Equal fluxes  for g=2}

 For this case the secular equation is simplified and takes the form of $ 4 [\Delta^{(even)} (\hat{\varphi}, \lambda) \Delta^{(odd)} (\hat{\varphi}, \lambda)]=1$.

For $N_{s}=2$, we solve analytically the secular equation. We find that the eigenvalues are given by $\lambda(n,\varphi;N=2)=r(N=2)\epsilon(n,\varphi;N=2)$, where $\epsilon(n, \varphi,N=2) = - 2 t \cos [ \frac{2 \pi}{N=2} (n + \hat{\varphi})]$, and $n=0, 1$ are the single ring eigenvalues. The value for $r(N=2)$ is $r(N=2)=\frac{\sqrt{3}}{2}$. To find the eigenvalues for other number of sites, $N_{s}= 6, 10, 14, 18, 22, 26, 30$, we numerically find the relation, $\lambda(n,\varphi;N)=r(N)\epsilon(n, \varphi;N)$, where $n = 0, 1, \ldots, N-1$ and $\epsilon(n, \varphi;N)= - 2 t \cos [ \frac{2 \pi}{N} (n + \hat{\varphi})]$ are the single ring eigenvalues. The function $r(N)$ is given in $figure$ $1a$. This figure shows that the function $r(N)$ reaches $one$ for large $N$. Using the function $r(N)$ given in figure $1a$, we compute the current for the $g=2$ case as a function of temperature,
$I^{(g=2)}(\varphi;N;T)=-\sum_{n=0}^{n=N-1}\frac{d} {d\varphi}[r(N)\cdot\epsilon(n, \varphi;N)]F(\frac{r(N)\cdot\epsilon(n, \varphi;N)-E_{fermi})}{K_{Boltzman}T})$.
 
$Figure$ $1b$ represents the current  for $N_{s}=30$ sites at two temperatures $T=0.02$ and $T=20$ $Kelvin$. In this figure, the current is given in dimensionless units $I$ plotted as a function of the dimensionless flux $f\equiv\hat{\varphi}_\alpha=[-0.5,0.5]$ ($\varphi_\alpha = 2 \pi \hat{\varphi}_\alpha = [-\pi,\pi]$). The solid line represents the single ring current and the dashed line represents the current for the genus $g=2$ case. In figure $1b$, the ratio of the currents at $T=0.02$ $Kelvin$ is $r(N=30,T=0.02)=0.979$.   
  
$Figure$ $1c$ shows that the currents at $T = 20$ Kelvin,  are in the range of $7$ $nA$ and the reduction of the current is larger in comparison with the $T = 0.02$ Kelvin case given in figure $1b$.

\textit{b) Two coupled rings with opposite fluxes , i.e. $\hat{\varphi}_1 = - \hat{\varphi}_2$}

For $N_{s}=2$, the eigenvalues are the $same$ as the one obtained for the same flux case. For $N_{s}= 6, 10, 14, \ldots, 2m+2$, we solve the secular equation given in eq. \ref{eq:secular} and compute the eigenvalues. In $figure$ $2a$, we plot the $total$ $energy$ as a function of the opposite fluxes at $T=0.02K$ for $30$ sites, $E^{(g=2)}(-\hat{\varphi},\hat{\varphi}, N_{s} = 30, T=0.02, K) = \sum_{n=0}^{n=N-1} [\lambda(-\hat{\varphi}, \hat{\varphi}) F(\frac{(\lambda(-\hat{\varphi},\hat{\varphi})-E_{fermi})}{K_{Boltzman}T})]$ . The total $energy$ dependence on the  $opposite$ $flux$ is  $chaotic$  due to    the interference between paths which $encircle$ zero and non zero fluxes, caused to the common point between the rings which acts as an impurity.  In addition we observe periodic oscillation with the $fundamental$ period $\Phi_{0}$ (see figure $2a$  and $2b$). For comparison, we show in $figure$ $2b$ the total energy for $equal$ fluxes which   is parabolic  and the current is linear (for  small fluxes).

 The effect of a finite width in the ballistic regime will give rise to a multichannel Persistent current which can be analyzed using the discussions for the multichannel case presented in the next chapter. Therefore the $energy$ will be given by a sum of energies   with respect the different channels. Therefore we  expect that the multichannel effect will give rise to a smooth function of  the  $energy$ as a function of  $opposite$ $flux$ .

\vspace{0.2 in}





\vspace{0.2 in}

\textbf{ IV  Discussion}

\textbf{ a) Possible experimental application of our  results.}
 
\vspace{0.2 in}

At this stage it is not clear if a genus g=2 experiments exists in the ballistic regime where our theory can be applied.
The  closest experiment which might be relevant  to our theory is the experiment presented in ref.13. The author of ref. 13 make the statement (see the last sentence in their paper) that from the theoretical side a model for the ballistic regime is needed for a direct comparison with the experiment. They make this statement  based on their results which indicate  that the  ratio  between the single ring current and the current in 16 coupled rings is close to one!  The authors of ref.13 say that the conditions in their experiment  are not in a clear diffusive  regions  therefore we can try to \textit{apply our ballistic theory to explain the experiments.}

The results presented  in  ref.[13]  are based on a system of    16 $GaAs/GaAlAs$ coupled rings.  At this stage  we have only results for two rings:
 We find for two rings the ratio $I^{(g=2)} / I_{single-ring}= 0.987$. Since the experiment was performed on 16 rings we use a scaling argument in order to extrapolate the results to 16 rings.  For two rings plus a $scaling$ argument we obtain, $r = I_{16-rings} / I_{single-ring} \approx [I^{(g=2)} / I_{single-ring}]^{4} = [r(T=0.02,N_{s}=50)]^{4} = [0.987]^{4} = 0.95$. \textit{The value $r=0.95$, is in the range of the experimental observation  reported [13]. }In the experiment the rings are connected trough arms of length of the order of the wavelength which can be approximated by point contact between the rings.

Next we  evaluate the amplitude of the current. Since at low temperature and large number of sites we have found that the ratio is close to one we will consider  the amplitude for a single ring. Using the experimental values 
for the  Fermi velocity $v_{F}=3.16 \cdot 10^{5}\cdot  m s^{-1}$,  Fermi wavelength   $\lambda_{F}=3.5\cdot 10^{-8} m $  and  the ring perimeter $L=1.2\cdot 10^{-5}m $  we  compute the persistent current for  a single ring. \textit{Using a model of a single  conducting  channel we  find at that the amplitude current at  $T=0$ is given by, $\frac{e v_{F}}{L}=4.2 n A$}. 
Using the  velocity  and  Fermi  wavelength we obtain the effective hopping constant used in our simulations, $t=\frac{\hbar v_{F}}{2 a \sin(K_{F} a)}$, using $K_{F}=\frac{2 \pi}{\lambda_{F}}$ with the effective lattice constant $a=\lambda_{F}/4$. This gives us that the effective length used in the simulation for 30 sites was $L_{simulation}=30\cdot\lambda_{F}/4=2.6\cdot10^{-7} m$.
Our simulation show that the current for 30 sites at $T=0.04$ K was $I=10^{-3}$ in dimensionless units which corresponds to a current $I\rightarrow I\cdot 0.92\cdot10^{-4} A$.
Using the value of the hopping constant $t=1.9 Joule=0.012 e.v.$ we find for $N_{s}$ a current ,
$I(N_{s}=30)=\frac{e}{\hbar}\frac{2 t}{N_{s}}=\frac{2.8}{N_{s}=30}\cdot 10^{-6}=92 nA$

In figure 3 we show the current dependence on the length of the ring for single and double rings at different temperature. At  T=0.02 K  the current decreases linearly with the length.
\textit{This allows to use  at T=0 the  linear relation between  $I(L=1.2\cdot 10^{-5} m)=4.2nA$ and $I(L_{simulation}=2.6\cdot 10^{-7} m)=92 nA$ which obey  the relation,  $\frac{I(L_{simulation})}{I(L)}=\frac {L}{L_{simulation}}$.}

\vspace{0.2 in}

\textbf{b) The effect of disorder on the Persistent current.}

\vspace{0.2 in}

We  explain the discrepancy of a factor of 10 between the experimental result and the theoretical calculation using  the $2K_{F}$impurity scattering for  a multichannel one dimensional rings at zero temperature.

 Contrarily to the transport current the Persistent current is without dissipation.
 The persistent current is determined by   the time derivative of the zero mode  coordinate (the macroscopic phase of the wave function). 
  In ref.9 (Schmeltzer  and Berkovits) it was shown that the effect of a  $2K_{F}$ impurity scattering gives rise to an enhancement of the kinetic mass and as a result the Persistent current is suppressed. \textit{Using the results given  in figure 1  (ref.9 Schmeltzer and Berkovits) we 
find that the suppression factor for one channel was $1-b W^2$, where $b=38.29$ and $W$is the $2K_{F}$ impurity scattering}. The value of $W$ can be obtained from the transport value 
 $l_{e}$, $l_{e}=8\cdot10^{-6} m$. From the Boltzman equation we will extract  $W$, 
 $\frac{v_{F} }{ l_{e}}=\frac{2\pi}{\hbar}\frac{t^{2}}{E_{F}}(\frac{W}{t})^{2}<(1- Cos(\theta))>$. The angular average is determined by the ratio of the width and length of the ring. We have,  $<1- Cos(\theta)>=\int_{-\theta_{0}}^{\theta_{0}}(1- Cos(\theta))\frac{d \theta }{2\pi}+\int_ {\pi-\theta_{0}}^{\pi+\theta_{0}}(1- Cos(\theta))\frac{d \theta }{2\pi}$ with $\theta_{0}=\frac{d}{L_{eff}}$  \textit{where $d=10^{-6} m$ is  the width of the ring } (which is actually  a square  with the length   $L_{eff}=\frac{L}{4}$) . Solving this equation we find that scattering  potential is given by,
$\frac{W}{t}= (\frac{\lambda_{F}}{l_{e} }\frac{d}{L}\frac{\pi}{32})^{1/2}=0.16$.
 
Using the result given in ref.9 (Schmeltzer and Berkovits) we compute the Persistent current for a single ring (with one channel) in the presence of the $2K_{F}$ potential  $\frac{W}{t}=0.16$. \textit{We find $\frac{e v_{F}}{L}(1-b W^2)=4.2 nA \cdot(1-38.26\cdot( 0.16)^2)=
0.09 nA$ which is smaller than the current observed in the experiment, $0.4 nA$.}
This discrepancy  suggest the possibility of a multichannel effect.
For a single ring the wave function is given by,  $\Psi(x,y)=\sum_{n=1}^{n_{max}} \psi_{n}(x)\chi_{n}(y)$ where $\chi_{n}(y)=(\frac{2}{d})^{1/2} Sin(n\frac{ \pi}{d})$ is the
standing wave in the transversal direction of the ring $0\leq y\leq d$. In the absence of disorder the problem is replaced by $n$ independent one dimensional channels with a shifted energy $\frac{\hbar^2}{2 m}(n\frac{ \pi}{d})^2$ and operators $\psi_{n}(x)$ ,$ \psi^\dagger_{n}(x)$. As a result the current will be given by, $I (no disorder)= \frac{e v_{F}}{L}\sum_{n=1}^{n_{max}}(1-(n\frac{\pi}{d}\frac{\lambda_{F}}{2\pi})^2)^{1/2}$.
The effect of the $2K_{F}$ impurity scattering will replace the current by,

$I= \frac{e v_{F}}{L}\sum_{n=1}^{n_{max}}(1-(n\frac{\pi}{d}\frac{\lambda_{F}}{2\pi})^2)^{1/2}[1-b W^{2}(1-(n\frac{\pi}{d}\frac{\lambda_{F}}{2\pi})^2)^{-1}]$  ( we have replaced $b$ by $b(1-(n\frac{\pi}{d}\frac{\lambda_{F}}{2\pi})^2)^{-1}$, in agreement with ref.9 where it was shown that $b$ is a function of the Fermi velocity for each  channel).
\textit{ Using the extracted values of $b$ and $W$ we find from the  condition $[1-b W^{2}(1-(n\frac{\pi}{d}\frac{\lambda_{F}}{2\pi})^2)^{-1}]\geq 0$ that the maximum number of conducting  channels  is given by, $n_{max}=8$. Using the value of $n_{max}=8$ we compute the current and find that the Persistent current is given   by  $I=0.4 nA$ which is in agreement with the experiment in ref.13.}

\textbf{the constraints for  two rings  which have a finite width.}

For this  case we have to do the  gluing  for a ring  of width $d$.
The transversal direction is $y$ and $x$is the direction of the one dimensional cchannel.
The constraints used previously are modified in the following way:
  $Q(x=0,y=d+\epsilon) = C_1(L/2,y=d+\epsilon') - C_2(0,y=d+\epsilon '')$ and $Q^+(0,y=d+\epsilon) = C^+_1(L/2,y=d+\epsilon' ) - C^+_2(0,y=d+\epsilon '')$.

Where $\epsilon'+\epsilon ''=l_{arm}$  and $l_{arm}=d$  represents the length of the arm which connects the   two rings.  As a result the common part between the rings will be given by the  connecting arm $l_{arm}=d$.

The second class constraints is given by,
 
$[Q(x\approx 0,y=d+\epsilon'),Q^+(x'\approx 0,y=d+\epsilon'')]_{+}
 =2 \delta(\epsilon'+\epsilon''-l_{arm})$

\textit{Using the second class constraint we obtain the equation of motion for the two rings   spinor $C_{\alpha=1,2}(x,y)$ }

$i\hbar \frac{d C_{\alpha}(x,y)}{d t} = [ C_{\alpha}(x,y), H_{T}]_D=
 [C_{\alpha}(x,y), H]$ 
 
$ - \int\,dz\int\,dz'[C_{\alpha}(x,y), Q^+(0,y=d+z]_+ ([Q^+(0,y=d+z), Q(0,y=d+z']_+)^{-1} [Q(0,y=d/2+z'), H]$
 
  $- \int\,dz\int\,dz'[C_{\alpha}(x,y), Q(0,y=d+z]_+ ([Q(0,y=d+z), Q^+(0,y=d+z')]_+)^{-1} [Q^+(0,y=d+z'), H]$
  
In the absence of disorder we can simplify the expression for the constraints if we introduce a contact point $y=d_{0}$  chosen   such  that,$d<d_{0}<l_{arm})$.  As a result we obtain a  constraints  for each channel, $Q_{n}(x=0)$ and $Q_{m}(x=0)$. We have the conditions   for  the second class constraints,
$[Q_{n},Q^{+}_{m}]_{+}=2\delta_{(n,m)}$. As a result the current for two rings will be similar as the one dimensional case . The  difference between the current for different channels in two rings will be determined by the  transversal shift  potential energy, $\frac{\hbar^2}{2 m}(n\frac{ \pi}{d})^2$.  We have repeated our simulation  for different transversal  potentials  with $n<n_{max}=8$  (the number of propagating  zero modes ) and find that the ratio between the double and single rings remains one.

\vspace{0.3 in}

\textbf{c) Generalization to many  coupled rings }

In order be able to study a system of many coupled rings we have to find a method which can be applied to many coupled rings . It seems that it might be preferable to replace the constraint of the single particle operator with the constraints defined in terms of the  electronic densities and currents. This method can be combined with the basic technique of matching  the boundary  conditions  used for  for solving  Quantum wires problems.  

 At the common 
point of the two rings at x=0  the constraint should gives rise  to equal densities ,$\rho_{1} \left( x=0\right)=\rho_{2} \left( x=0\right)$.  Formally this condition 
is enforced  with the help of the scalar   field  $ a_{0}\delta \left( x\right)$ which plays the role of the Lagrange multiplier.  
For g-coupled rings  we introduce a set  
 of scalar  potentials $ a^{( 0,1)}_{0}
...,..a^{( g-2,g-1)}_{0} $ for which a statistical
$annealed$ average  has to be performed.  

The hamiltonian for $"g"$ coupled
rings with $\rho _{n}\left( x\right) $ electronic density and  $\varphi _{1}.....\varphi _{g}$ fluxes ,

\vspace{0.2 in}

$H^{\left( g\right) }=H\left( \varphi _{1}...\varphi _{g}\right)
+\sum_{n=0}^{g-2}a_{0}^{\left( n,n+1\right) }\left( \rho _{n}\left( x-\left(
n+1\right) L\right) -\rho _{n+1}\left( x-nL\right) \right)$

To demonstrate  this method  we repeat the case    g=2  two ring case.

For  a symmetric
configuration with the common point  at $x=0$, we fold the
space of the first ring from $\left[ -L,0\right] $ to $\left[ L,0\right] $
such that the space of the two ring is restricted to $0\leq x\leq L$.  The wave function for the coupled rings   $Z_{E}(x)$ with  eigenvalue $E$  is given as a $spinor$ with  two components $Z_{\tau }\left( x\right) $, $\tau=1,2$. The  Schr\"{o}dinger equation for the two rings in the presence of the  field  $ a_{0}\delta \left( x\right)$ is,
\begin{equation}
\left[ -\left( -\partial _{x}-i\frac{2\pi }{L}\varphi _{1}\right) ^{2}+%
{a}_{0}\delta \left( x\right) \right] Z_{1}\left( x\right) 
=K^{2}Z_{1}\left( x\right)\nonumber\\
\end{equation}

\begin{equation}
\left[ -\left( \partial _{x}-i\frac{2\pi }{L}\varphi _{2}\right) ^{2}-
a_{0}\delta \left( x\right) \right] Z_{2}\left( x\right) 
=K^{2}Z_{2}\left( x\right)\nonumber\\
\end{equation}
 Where $K^{2}\equiv \frac{2m}{\hbar ^{2}}E$  with the    energy $E$. The  current $I\left[ \varphi _{1},\varphi _{2};a_{0}%
\right] $ is a function of $\varphi _{1}$, $\varphi
_{2}$ and $a_{0}$.

The eigenfunctions  for this problem are obtained by matching  the boundary conditions:
\textit{a)the continuity  of the wave functions at the boundary  x=0 ,} $Z_{1}\left( x\right) =Z_{1}\left( x+L\right) $ and $\ Z_{2}\left( x\right)
=Z_{2}\left( x+L\right) $ 

b)  \textit{The discontinuity of the derivative at x=0 for each ring,}

 $\left( -\partial _{x}-i\frac{2\pi }{L}\varphi _{1}\right) Z_{1}\left(
x=-\varepsilon \right) -\left( -\partial _{x}-i\frac{2\pi }{L}\varphi
_{1}\right) Z_{1}\left( x=\varepsilon \right) 
=a_{0}Z_{1}\left( x=0\right)$ 

$\left( \partial _{x}-i\frac{2\pi }{L}\varphi _{2}\right) Z_{2}\left(
x=-\varepsilon \right) -\left( -\partial _{x}-i\frac{2\pi }{L}\varphi
_{2}\right) Z_{2}\left( x=\varepsilon \right)   
=-a_{0}Z_{2}\left( x=0\right)$

(The change of sign of the flux in  ring  one $\alpha=1$  is a result of folding the space  from $[-L,0]$ to $[L,0]$.)

As a result the  current $I\left[ \varphi _{1},\varphi _{2};a_{0}%
\right] $ is a function of $\varphi _{1}$, $\varphi
_{2}$ and  the scalar potential $a_{0}$.  \textit{The physical current will be  obtained  after  averaging  over the constraint  $a_{0}$, $\bar{I}\left( \varphi
_{1},\varphi _{2}\right) =\lim\limits_{\Lambda\rightarrow \infty }\int_{-\Lambda}^{\Lambda}%
\frac{d a_{0}}{2\Lambda}I\left( \varphi _{1},\varphi _{2};a
_{0}\right) $.}
The need for the additional average makes the solution  for the persistent current  more involved. 
 \textit{ Therefore  the resulting  method is  more complicated  in comparison with $Dirac's$ method used in the first part for two rings. From other hand for  for many rings
 the new method seems to be preferable to the $Dirac's$ method. } This becomes clear  once the
electron-electron interactions  and disordered are considered, this new method allows the use of  field theory   methods such as  BOSONIZATION  and Renormalization Group [9,10].

\textbf{Summary}

In this paper, we have introduced a method which solves the problem of the $global$ phase of the wave function for geometrical structures with holes, i.e. high genus materials. This method is applicable to a variety of mesoscopic systems where coherency of wave function is important.

We have found an exact solution for the persistent current in two coupled rings. By numerical calculations, we have computed the current dependence on the flux, temperature, and the number of sites.
This theory might be tested in  coupled rings  for equal and opposite flux in the Ballistic regime. 
A possible explanation of the experiment in ref.13 has been proposed.

\begin{bibliography}{99}
1. Y. Aharonov, D.Bohm, Phys. Rev. 115, 485 (1959). \\
2. N. Byers and C.N. Yang, Phys. Rev. Lett. 7, 46 (1961). \\
3. M. Buttiker, Y. Imry, and R. Landauer, Phys. Lett. A. 96, 365 (1983).\\
4. L.P. Levy, G. Dolan, J. Dunsmuir, and H. Bouchiat, Phys. Rev. Lett. 64, 2074 (1990).\\
5. Y. Gefen, Y. Imry, and M.Y. Azbel, Phys. Rev. Lett. 52, 129 (1984). \\
6. P.Singha Deo,Phys.Rev.B 53,15447 (1996).\\
7. T.P.Pareek and A.M. Jayannavar,Phys.Rev. 54,6376 (1996).\\
8. H. Aoki, J. Phys. C. 18, 1885-1890 (1981).\\
9. D. Schmeltzer, Phys. Rev. B. 63, 125332 (2001); D. Schmeltzer and R. Berkovits, Physics Letters A 253, 341 (1999).\\
10. C.L. Kane and M.P.A. Fisher, Phys. Rev. Lett. 68, 1220 (1992).\\
11. Ken-Ichi Sasaki and Yoshiyuki Kawazoe, Cond-Mat/0408505.\\
12. K. Sasaki, Y. Kewazoe, and R. Saito, Physics Letters A321, 369-375 (2004).\\
13. W. Rabaud, L. Saminadayar, D. Mailly, K. Hasselbach, A. Benoit, and B. Etienne, Phys. Rev. Lett. 86, 3124 (2001).\\
14. Paul A. M. Dirac, ``Lectures on Quantum Mechanics,'' Belfer Graduate School of Science, Yeshiva University, New York, 1964.
\end{bibliography}

\pagebreak

\begin{figure}
\includegraphics[width=4in]{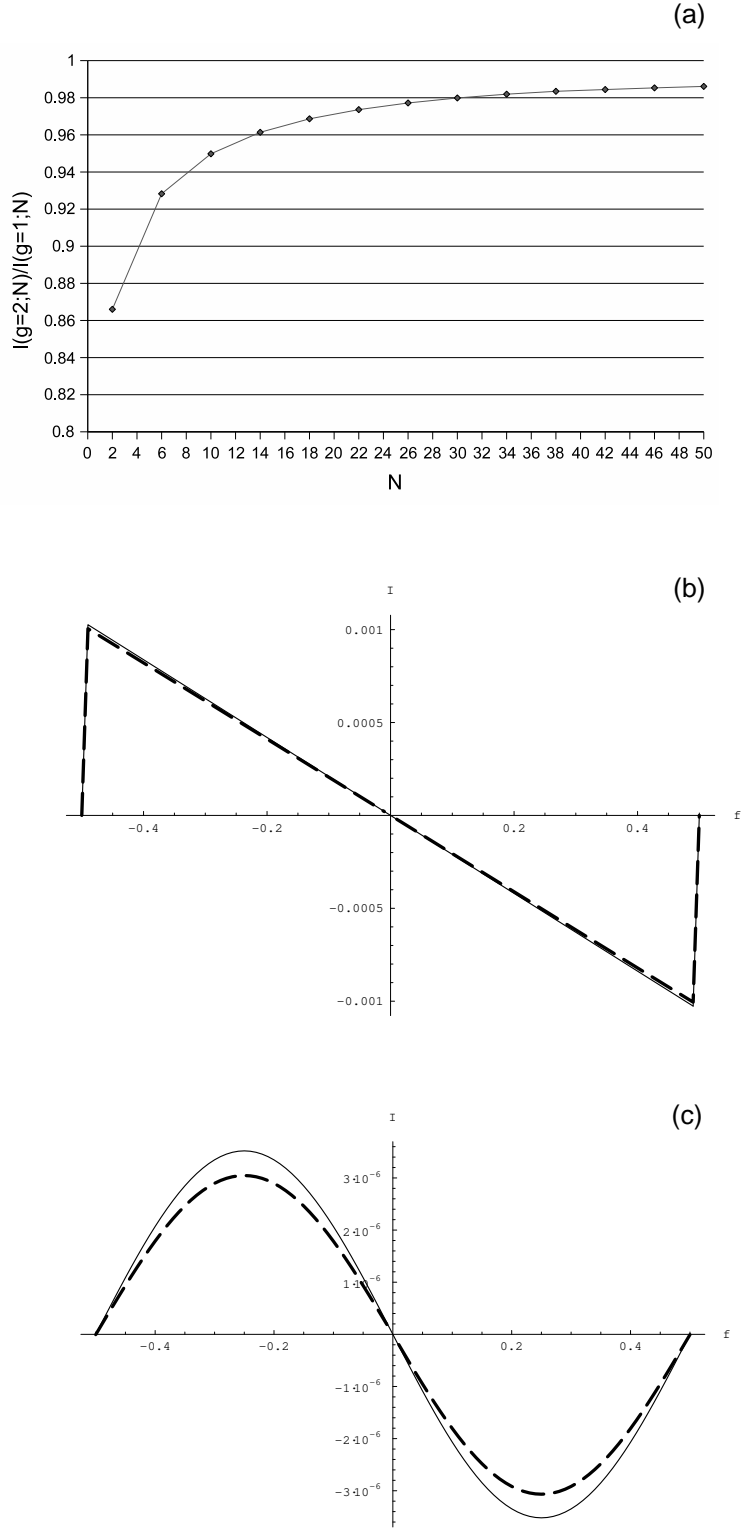}
\caption{(a) The ratio of the double to single ring currents $I(g=2;N)/I(g=1;N) = r(N)$; (b) The single ring (solid line) and the double ring (dashed line) currents for $N_{s}=30$ at $T = 0.02$ Kelvin; and (c) The single ring (solid line) and the double ring (dashed line) currents for $N_{s}=30$ at $T = 20.0$ Kelvin.}
\end{figure}

\begin{figure}
\includegraphics[width=4in]{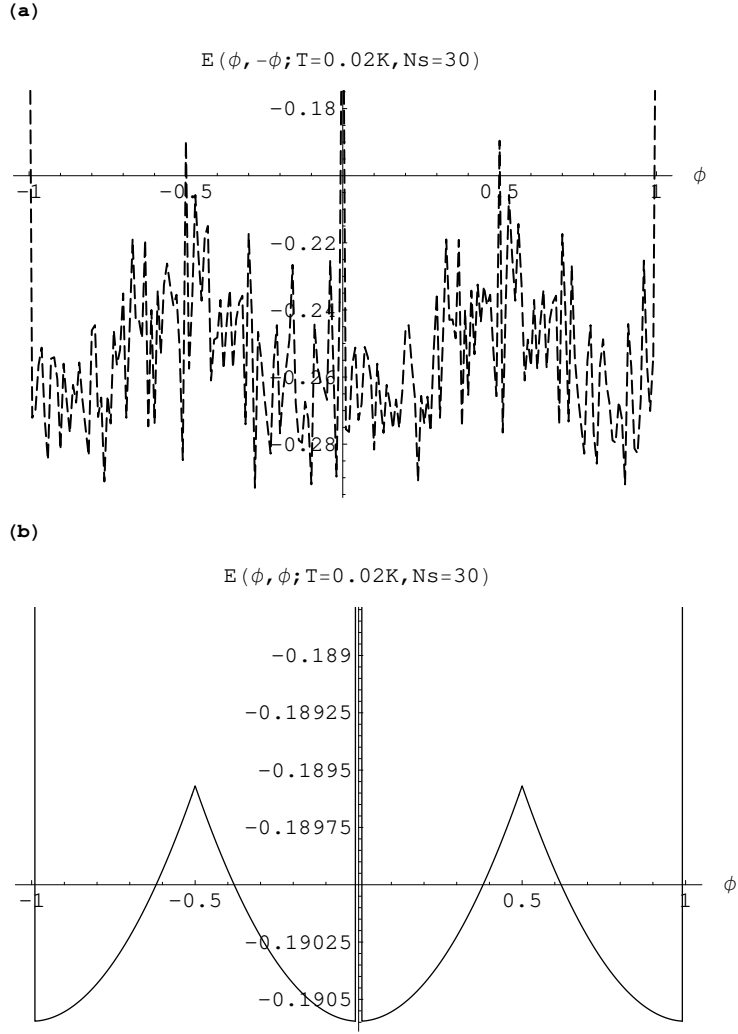}
\caption{(a) The total energy for $opposite$ fluxes, $\phi=\hat{\varphi_1}=-\hat{\varphi_2}$ for $30$ sites at $T=0.02$ Kelvin $E^{(g=2)}(-\phi,\phi;N_{s}=30,T=0.02 K)$ ;and (b) The total energy for $equal$ fluxes ,$\phi=\hat{\varphi_1}=\hat{\varphi_2}$ $E^{(g=2)}(\phi,\phi;N_{s}=30,T=0.02 K)$ }
\end{figure}

\begin{figure}
\includegraphics[width=4in]{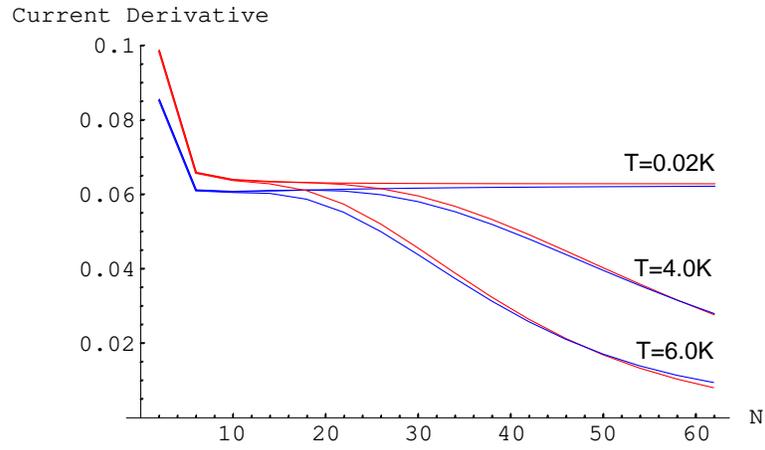}
\caption{The current derivative with respect the flux at $\phi<<1$ multiplied by the number of sites $N$ as a function of $N$ at different temperatures-$I^{'}\cdot N$ for a single and a double ring}
\end{figure}
 
\end{document}